\documentclass{PoS}

\title{QCD thermodynamics with $N_f=2+1$ near the continuum limit at realistic quark masses}

\ShortTitle{QCD thermodynamics with $N_f=2+1$}

\author{\speaker{Takashi Umeda} for the
        RBC-Bielefeld Collaboration\\
        Brookhaven National Laboratory, Upton, NY 11973, USA\\
        E-mail: \email{tumeda@bnl.gov}}

\abstract{We report on our study of QCD thermodynamics with $2+1$
flavors of dynamical quarks. 
In this proceeding we present several thermodynamic quantities
and our recent calculation of the critical temperature. 
In order to investigate the thermodynamic properties of QCD 
near the continuum limit we adopt
improved staggered (p4) quarks coupled with tree-level Symanzik
improved glue on $N_t=4$ and $6$ lattices. 
The simulations are performed with a physical value of the strange quark
mass and light quark masses which are in the range of
$m_q/m_s=0.05-0.4$. 
The lightest quark mass corresponds to a pion mass of about 150 MeV.}

\FullConference{XXIV International Symposium on Lattice Field Theory\\
		 July 23-28 2006\\
		 Tucson Arizona, US}

\begin{document}

\section{Introduction and Lattice Setup}
The calculation of QCD thermodynamics from first principle is
important for various research areas such as Heavy Ion Phenomenology,
Cosmology and Astrophysics. Lattice QCD enables us to carry out such
calculations. Especially for HIC phenomenology it is mandatory to improve
estimates on some basic thermodynamic quantities which have been obtained in
previous lattice calculations. 
One of the main subjects in our project is the accurate determination of the
critical temperature $T_c$, whose uncertainty, for example, strongly affects 
the critical energy density $\epsilon_c$ because of its $T_c$ dependence,
$\epsilon_c \sim T_c^4$.

Since thermodynamics of lattice QCD requires huge
computational resources, it is difficult to perform an ideal simulation.
Recent studies tell us that quark masses and the number of flavors strongly 
affect thermodynamic quantities \cite{Nf}. Reliable continuum extrapolations
are of tremendous importance as well \cite{a=0}.  
Therefore, it is our aim to study QCD thermodynamics with almost
realistic quark masses on the QCDOC machine at Brookhaven National
Laboratory and the APEnext machine at Bielefeld University.
The calculation is performed with $N_f=2+1$, which means 2 degenerate
light quarks and one heavier quark on lattices with $N_t=4$ and 6.
The lightest quark masses of our simulation yields a pion mass of about 150
MeV and a kaon mass of about 500 MeV.  

For such calculations we adopt the p4fat3 quark action, which is
an improved Staggered quark action \cite{p4fat3}, 
with a tree-level improved Symanzik gauge action. 
By using the p4fat3 action,
the free quark dispersion relation has the continuum form
up to $O(p^4)$, and the taste symmetry breaking is suppressed by a
3-link fattening term. 
The action also improves bulk thermodynamical quantities
in the high temperature limit \cite{p4fat3}.
The improvements are essential to control the continuum extrapolation 
on rather coarse lattices, i.e. $N_t=4$ and 6. 
The gauge ensembles are generated by an exact RHMC algorithm \cite{RHMC}.

As a status report of the project, in this proceeding, we present
several thermodynamic quantities, which are order parameters and their
susceptibilities, the static quark potential, and the spatial string tension.
In the last section we discuss the critical temperature at the physical
point.
The details of the critical temperature calculation are given in our recent
paper \cite{Tc}.

\section{Order Parameters and Susceptibilities}
\label{sec:order}
To investigate the QCD critical temperature and phase diagram, order
parameters 
of the QCD transition are indispensable. 
In the chiral limit the chiral condensate $\langle
\bar{\psi}\psi \rangle$ is the order parameter for the spontaneous
chiral symmetry breaking of QCD. On the other hand in the heavy quark
limit the Polyakov loop $\langle L \rangle$ is the order parameter 
of the deconfinement phase transition.
For finite quark masses, these observables remain good
indicators for the (pseudo) critical point.
Especially their susceptibilities are useful to determine the critical
coupling $\beta_c$ in numerical simulations.

\begin{figure}
\begin{center}
 \includegraphics[width=.4\textwidth]{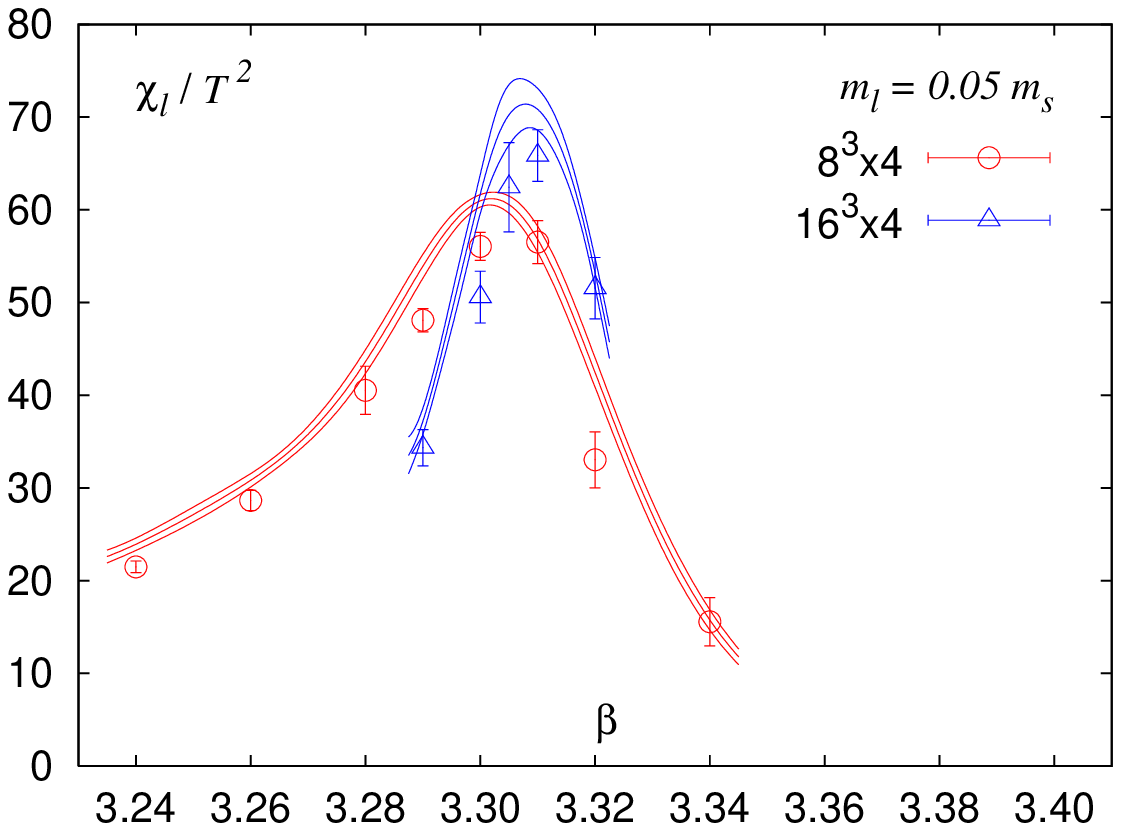}
 \includegraphics[width=.4\textwidth]{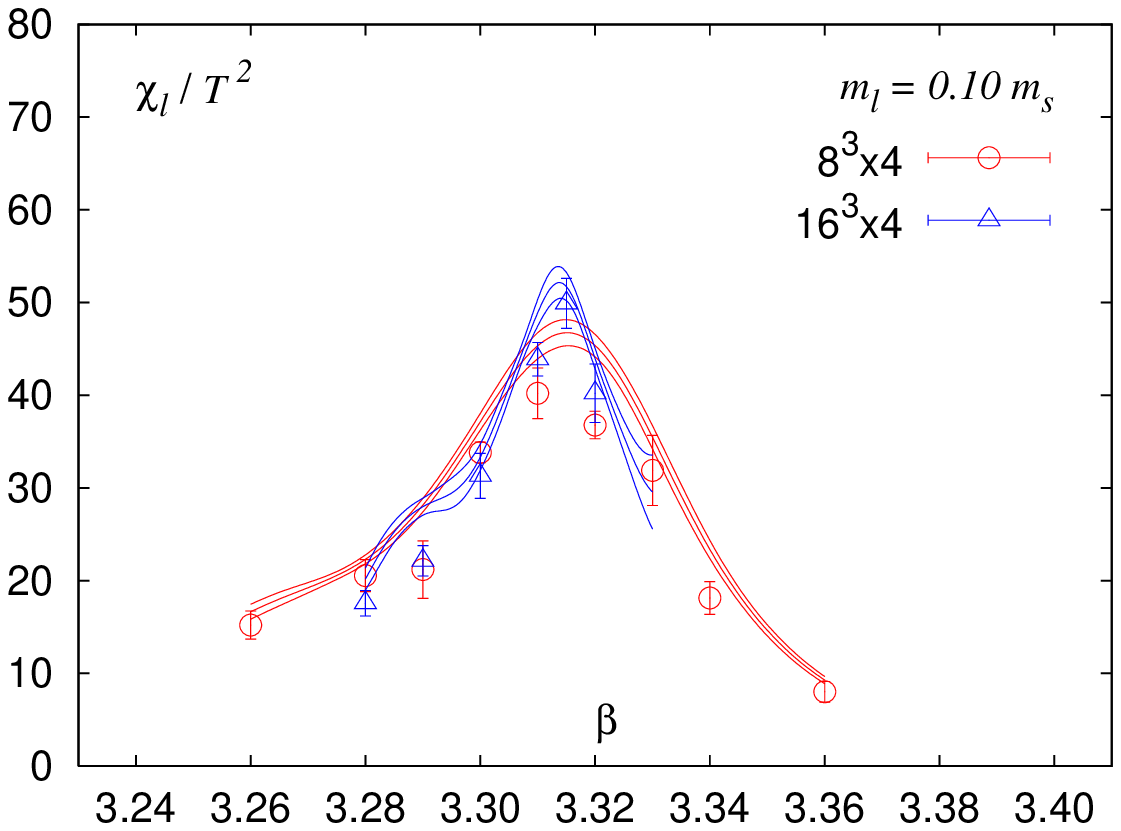}
 \includegraphics[width=.4\textwidth]{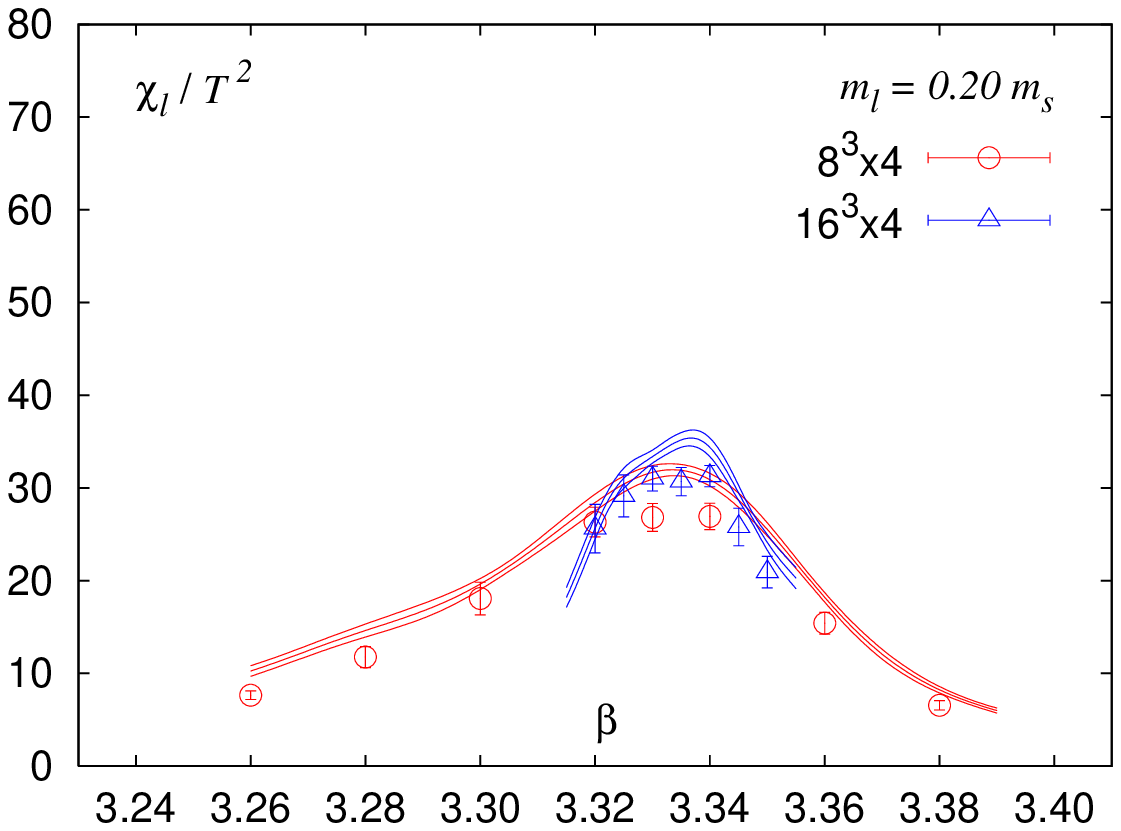}
 \includegraphics[width=.4\textwidth]{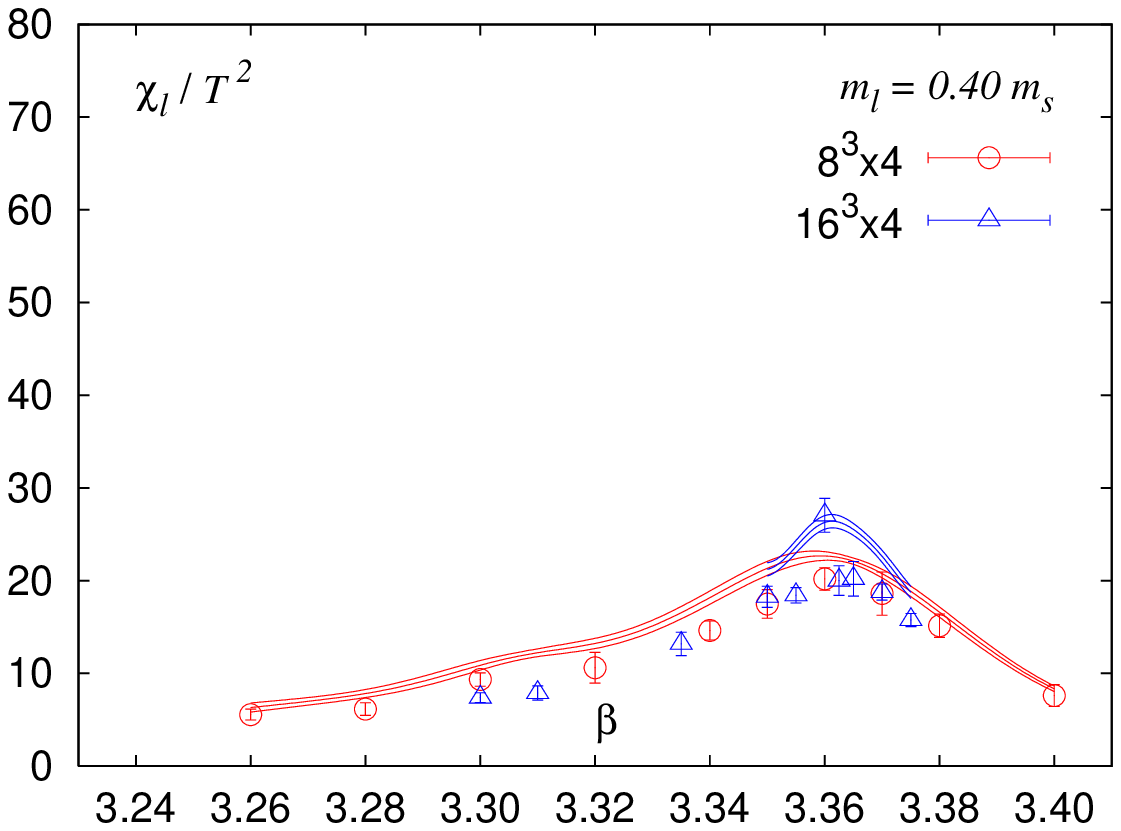}
\caption{The susceptibility of the light quark chiral condensate on
 $8^3\times 4$ and $16^3\times 4$ lattices. 
Each panels correspond to results with $\hat{m}_l/\hat{m}_s = 0.05, 0.1, 0.2$ 
and $0.4$ respectively.
The lines are calculated by the multi-histogram re-weighting technique.}
\label{fig_suscep}
\end{center}
\end{figure}

Figure~\ref{fig_suscep} shows the susceptibilities of the light quark
chiral condensate \footnote{We always use here the disconnedted part of
the chiral susceptibility.}
 on $8^3 \times 4$ and $16^3\times 4$ lattices with 
various quark masses. The peak positions of the susceptibilities 
define the point of the most drastic change of each order parameter, 
i.e. the (pseudo) critical point of the QCD transition. 
The results are interpolated in the coupling $\beta$ by using the 
multi-histogram re-weighting technique \cite{multi}.

The strength of the transition decreases with increasing quark masses,
this is reflected in the decreasing peak height of the chiral susceptibilities.
We calculate these susceptibilities on lattices with aspect ratios of
$N_s/N_t=2$ and 4. Since we see a rather small volume
dependence the results suggest that the transition is in fact not a true
phase transition in the thermodynamic sense but a rapid crossover. 

\begin{figure}
\begin{center}
\includegraphics[width=.5\textwidth]{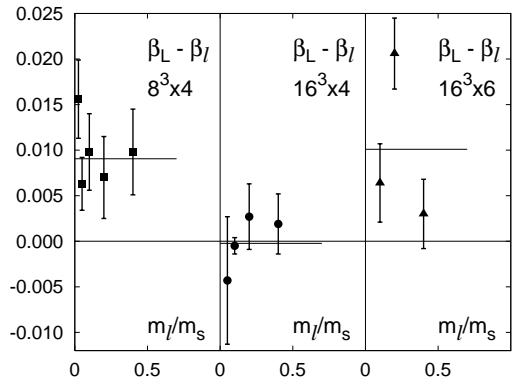}
\caption{The difference of gauge couplings at the location of peaks in
 the Polyakov loop and the chiral susceptibilities, $\beta_L-\beta_l$.}
\label{beta_c}
\end{center}
\end{figure}

Figure \ref{beta_c} (left) shows the difference in 
the peak position of the chiral and Polyakov loop susceptibilities.
The differences are small and almost identical at the aspect ratio of
$N_s/N_t=4$,  which indicates 
the chiral and deconfinement phase transition occur at almost
the same temperature. The discrepancy between the peak positions
shrinks with increasing volume.

\section{Scale Setting and the Heavy Quark Potential}
The lattice scale is determined by the heavy quark potential $V(r)$
which is extracted from Wilson loops.
The Wilson loop expectation values are calculated on $16^3 \times 32$ 
lattices with APE smearing in spatial direction.
The spatial path in a loop is determined by the Bresenham algorithm \cite{BA}.
We calculate the string tension, $\sigma$
and Sommer scale $r_0$, which is defined \cite{r0} as the distance where the 
corresponding force of the static quark potential
matches a certain value suggested by phenomenology:
$r^2\frac{\partial V}{\partial r}\mid_{r=r_0}=1.65$. 
To remove short range lattice artifacts we use the improved
distance, $r_{imp}$, which is defined as
\begin{equation}
\frac{1}{4\pi r_{imp}}\equiv \int\frac{d^3k}{(2\pi)^3}
\frac{e^{ikr}}{4\sum_i(\sin^2{\frac{k_i}{2}})
+\frac{1}{3}\sin^4{\frac{k_i}{2}}}.
\end{equation}

\begin{figure}
\begin{center}
 \includegraphics[width=.4\textwidth]{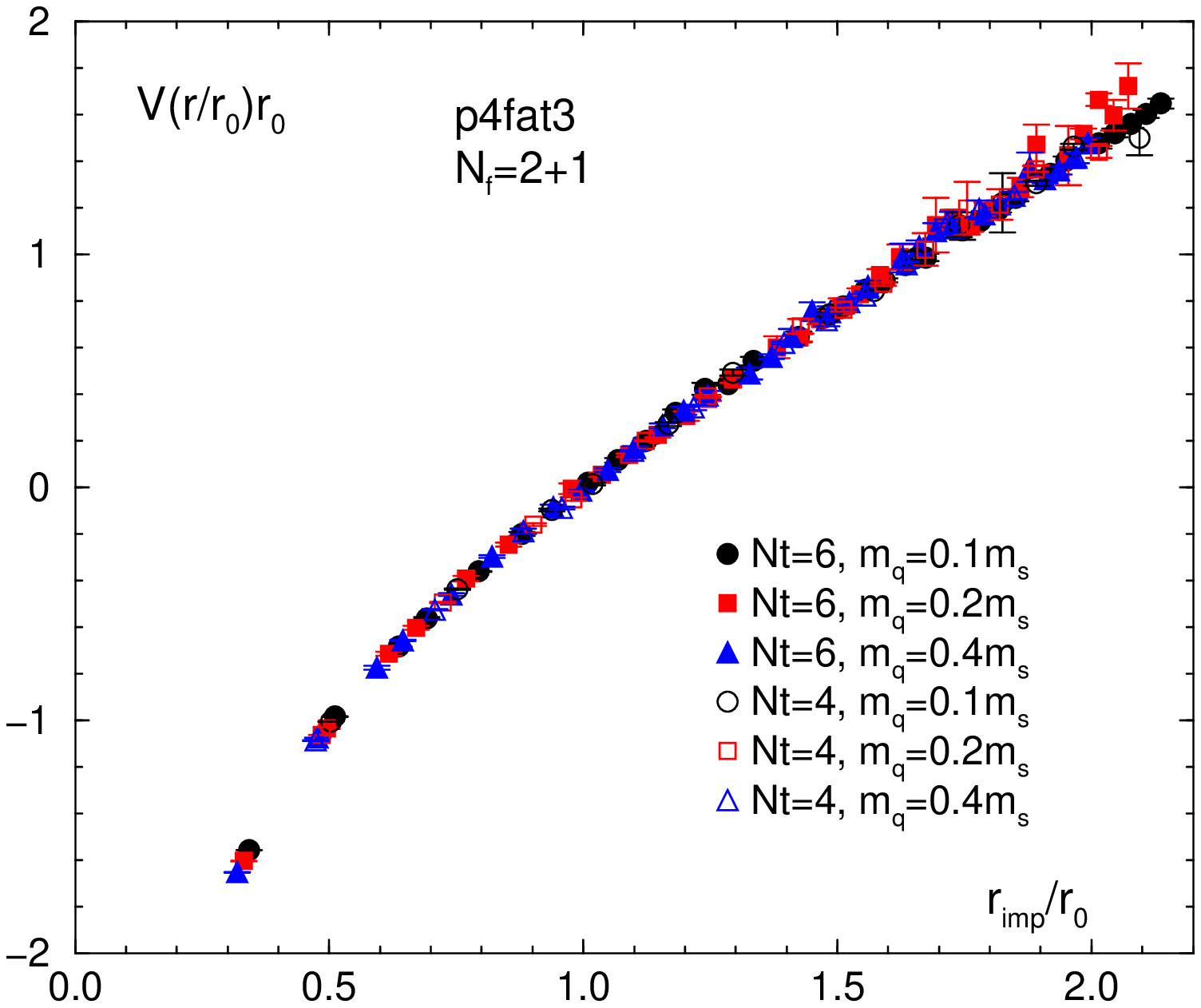}\hspace{5mm}
 \includegraphics[width=.41\textwidth]{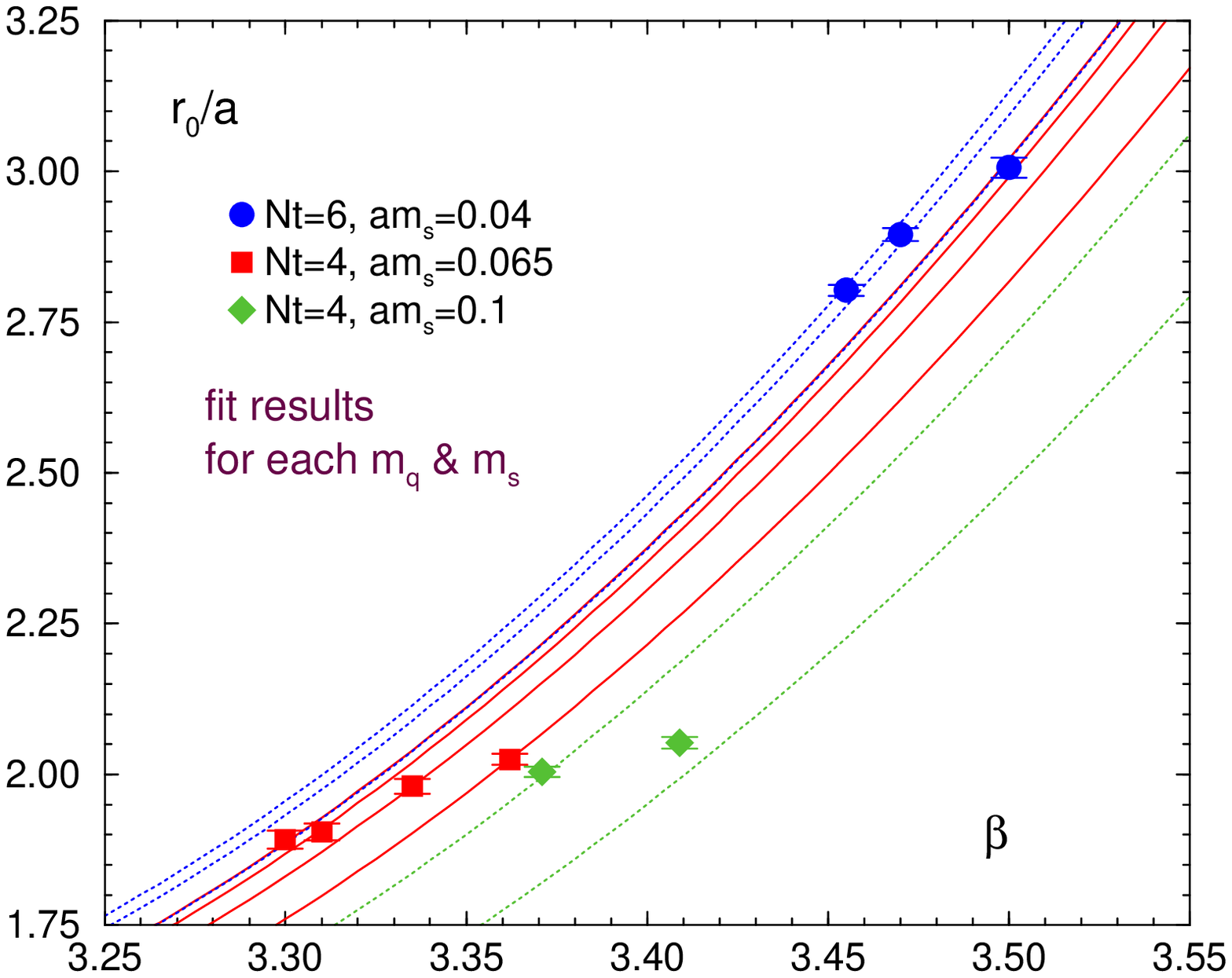}
\caption{Heavy quark potentials scaled by $r_0$ for various cut-offs and 
quark masses (left) and a global fit of $r_0/a$ with the RG inspired
 ansatz Eq.(3.3) (right).}
\label{fig:r0}
\end{center}
\end{figure}

In our lattice setup, we find almost no mass and cutoff dependence in
the potential scaled by $r_0$ at $N_\tau=4$ and 6 (Fig.2(left)). 
As discussed in previous studies \cite{sb}, 
we also find no string breaking effects even at large $r$.
To estimate systematic uncertainties of the potential fit, we performed 
several types of fits, e.g. different fit-ranges in $r$ 
and fit-forms (3 \& 4 params. fits),
\begin{equation}
V(r)=C+\frac{\alpha}{r_{imp}}+\sigma r_{imp},
~~~
V(r)=C+\frac{\alpha}{r}+\sigma r+d(\frac{\alpha}{r_{imp}}-\frac{\alpha}{r}).
\end{equation}
The differences in the mean values of the fits are taken into account as
a systematic uncertainty of the scale setting. 

We have determined the scale parameter $r_0$ in units of the lattice
spacing for 9 different parameter sets. 
This allows to interpolate between different values of the gauge coupling
and quark masses.
We use a renormalization group inspired ansatz \cite{allton} which takes 
into account the quark mass dependence of $r_0/a$ 
\cite{Bernard04} and which approaches, in the weak coupling limit,
the 2-loop  $\beta$-function for three massless flavors,
\begin{equation}
(r_0/a)^{-1} = 
R(\beta) (1 +B \hat{a}^2(\beta) + C \hat{a}^4(\beta))  
{\rm e}^{A (2 \hat{m}_l + \hat{m}_s)+D} 
\; .
\label{interpolate}
\end{equation}
Here $R(\beta)$ denotes the 2-loop $\beta$-function  
and $\hat{a}(\beta) =R(\beta)/R(\bar{\beta})$ with $\bar{\beta}=3.4$ chosen
as an arbitrary normalization point.

\section{Spatial String Tension}

Let us now discuss the calculation of the spatial string tension which is
important to verify the theoretical concept of dimensional reduction at high
temperatures. 
The spatial string tension is extracted from the spatial static quark
``potential''  (from spatial Wilson loops).
We use the same analysis technique as for the usual (temporal) static quark
potential.

At high temperature, the spatial string tension $\sigma_s(T)$ is
expected to behave like 
\begin{eqnarray}
 \sqrt{\sigma_s(T)}&=&cg^2(T)T.\label{2loop}
\end{eqnarray}
Here $g^2(T)$ is the temperature dependent coupling constant from the 
2-loop RG equation,
\begin{eqnarray}
g^{-2}(T)&=&2b_0\ln{\frac{T}{\Lambda_\sigma}}+\frac{b_1}{b_0}
\ln{\left(2\ln{\frac{T}{\Lambda_\sigma}}\right)}.
\end{eqnarray}
If dimensional reduction works, the parameter``$c$'' should be equal to the
3-dimensional string tension and should be flavor independent.

\begin{figure}
\begin{center}
\includegraphics[width=.6\textwidth]{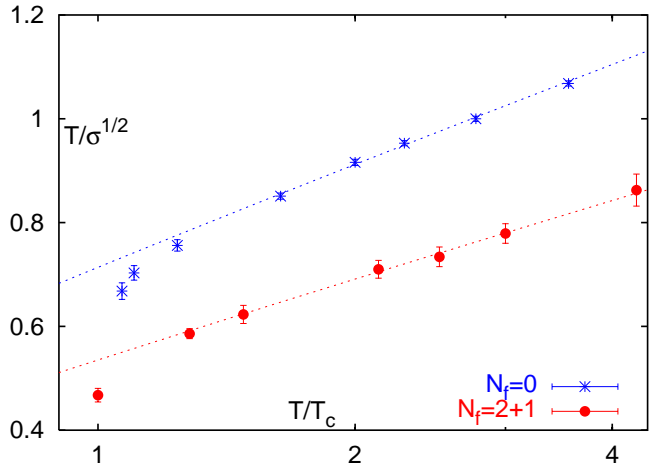}
\caption{Temperature dependence of the spatial string tension for
 $N_f=0$ and $N_f=2+1$. Both dotted lines are fits with
 Eq.~(4.1). Note, that the scale on the horizontal axis is
 logarithmic.}
\label{sigma_s}
\end{center}
\end{figure}

Our 2+1 flavor result yields $c=0.587(41)$ and $\Lambda_\sigma/T_c=0.114(27)$,
obtained by a fit with Eq.(\ref{2loop}). On the other hand, we plot in
Fig.~\ref{sigma_s} also the quenched result \cite{spst} which gives
$c=0.566(13)$ and $\Lambda_\sigma/T_c=0.104(9)$. We thus find that the 
parameter ``c'' is -- within statistical errors -- independent on the number
of dynamical flavors and that dimensional
reduction works well even for $T=2T_c$. 
This analysis can and will be refined in the future by taking into
account higher order corrections to Eq.(\ref{2loop}) \cite{higher}.

\section{The transition temperature}
Finally I discuss the transition temperature in QCD, which is one of the
most important subjects in our project.
In sect.\ref{sec:order} we have determined the critical $\beta$ at each
$N_\tau$ and for several quark masses.
At these couplings we performed zero temperature
calculations of the static quark potential.
The scale settings at each critical $\beta$ provide each
values for the critical temperature. 
We thus can extrapolate $T_c$ to the chiral
limit as well as to the physical point by using a scaling ansatz.
In the zero temperature calculations the actual $\beta$ is sometimes
slightly different from our final result on the critical $\beta$. The
differences are corrected by using Eq.\ref{interpolate}. 
Furthermore a systematic error in
the critical $\beta$, e.g. $\beta_L-\beta_l$ at $N_\tau=6$ in
Fig.\ref{beta_c}, is also taken into account in the critical
temperature. 

\begin{figure}
\begin{center}
 \includegraphics[width=.45\textwidth]{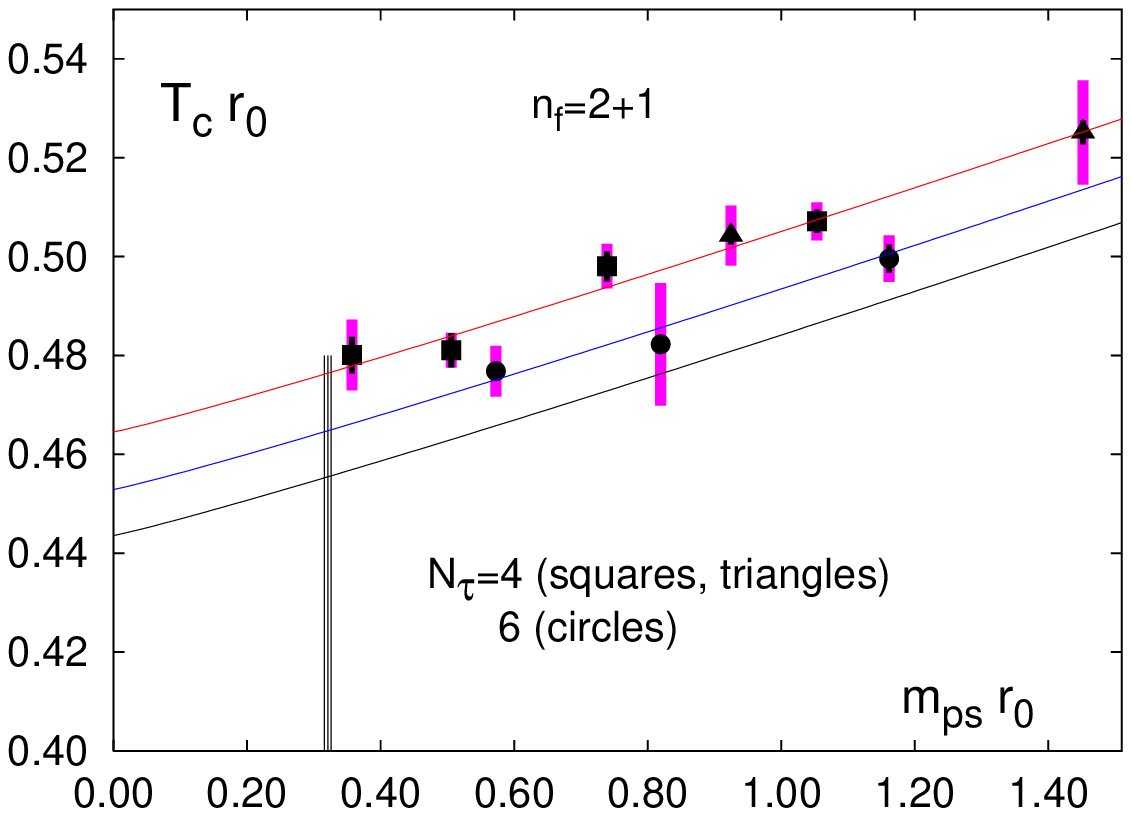}
 \includegraphics[width=.45\textwidth]{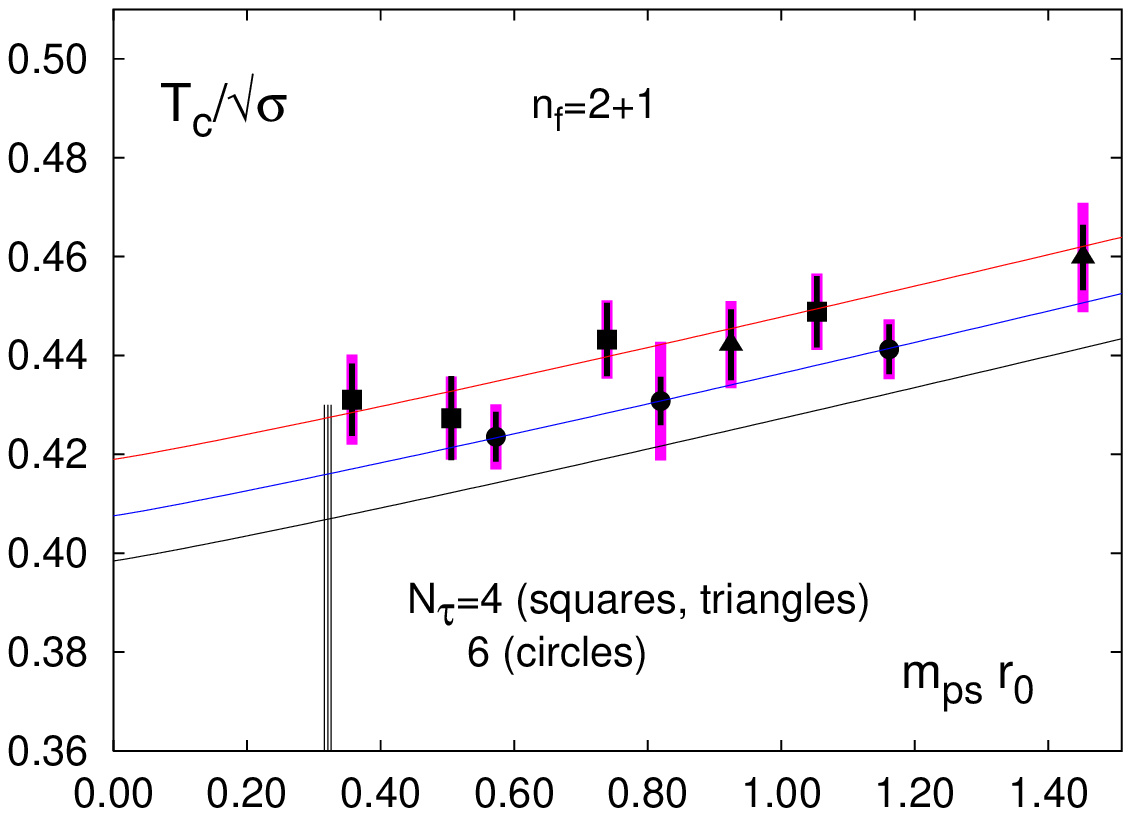}
\caption{$T_c r_0$ (left) and $T_c/\sqrt{\sigma}$ (right) as a function of 
$m_{ps}r_0$ on lattices with temporal extent $N_\tau =4$, $\hat{m}_s= 0.065$
(squares) and $\hat{m}_s= 0.1$ (triangles) as well as for 
$N_\tau = 6$, $\hat{m}_s= 0.04$ (circles).
Thin error bars represent the statistical and systematic error on
$r_0/a$ and $\sqrt{\sigma}a$. The broad error bar combines this error with
the error on $\beta_c$. 
The vertical line shows the location of the physical value 
$m_{ps} r_0= 0.321(5)$ and its width represents the error on $r_0$.
The three parallel lines show results of fits based on
 Eq.~(5.1) with $d=1.08$ for $N_\tau=4,~6$ and $N_\tau \rightarrow \infty$ 
(top to bottom).} 
\label{fig:tc}
\end{center}
\end{figure}

In Fig.~\ref{fig:tc} we plot the critical temperature in unit of the
Sommer scale (left) and the string tension (right) as function of the
pion mass $m_{PS}$ (also in units of the Sommer scale).
Thin error bars represent the statistical and systematic error on
$r_0/a$ and $\sqrt{\sigma}a$. The broad error bar combines this error with
the error on $\beta_c$. 
We perform a combined chiral and continuum extrapolation of $T_c$ by
using the ansatz, e.g. in unit of the Sommer scale,
\begin{equation}
T_c r_0= \left[T_cr_0\right]^{\rm chiral}_{\rm cont}+
A(m_\pi r_0)^d+B/N_\tau^2 ,
\label{eq:extrapolate}
\end{equation}
where $A$ and $B$ are free fit parameters. 
If the QCD transition is second order in the chiral limit the transition
temperature is expected to depend on the quark mass as
$\hat{m}_l^{1/\beta\delta}$, which corresponds to $d\simeq 1.08$ in
Eq.~\ref{eq:extrapolate}, by the fact that one expects a
critical point in the chiral limit which is in the $O(4)$-universality
class. 
If, on the other hand, the transition becomes first order for small
quark masses, the transition temperature will depend linearly on the
quark mass, i.e. $d=2$. 
Using the fit form ansatz \ref{eq:extrapolate} we can determine the
transition temperature at 
the physical point, which is defined as $m_{ps}r_0\equiv 0.321(5)$,
\begin{equation}
T_c r_0= 0.457(7)^{+12}_{-3},~~~T_c/\sqrt{\sigma}=0.408(8)^{+3}_{-1},
\label{eq:tcr0phys}
\end{equation}
where the central value is given for fits with $d=1.08$ and the lower
and upper systematic error correspond to $d=1$ and $d=2$, respectively.

In order to convert $T_c$ to physical units, we use the scale parameter,
$r_0=0.469(7)$ fm,  deduced from the bottomonium level splitting using
NRQCD \cite{NRQCD}. Finally we obtain the transition temperature in QCD
at the physical point, $T_c=192(7)(4)$ MeV, where the statistical error
includes the errors given in Eq.~\ref{eq:tcr0phys} 
as well as the uncertainty in the value of $r_0$ and the second error
reflects our estimate of a remaining systematic error on the
extrapolation to the continuum limit.
The value of the critical temperature obtained here is about 10\% larger 
than the frequently quoted value $\sim 175$~MeV.
We note that this larger value mainly results from  
the value for $r_0$ used in our conversion to physical scales.

The analysis presented here leads to a value for the critical temperature with 
about 5\% statistical and systematic errors. It clearly is desirable to 
confirm our estimate of the remaining systematic errors through an additional
calculation on an even finer lattice. Furthermore, it is desirable to 
verify this result through calculations using other $T=0$ scales and
to explore other discretization
schemes for the fermion sector of QCD and to also obtain a reliable 
independent scale setting for the transition temperature from an observable 
not related to properties of the static quark potential.

\acknowledgments

The simulation has been done on QCDOC at Brookhaven National Laboratory
and APEnext at Bielefeld University.
This work has been supported by the the U.S. Department of Energy under
contract DE-AC02-98CH1-886.

\end{document}